\documentclass{sig-alternate}
\usepackage{times,amsmath,epsfig,multirow,tabularx,supertabular}

\newcolumntype{Z}{>{\centering\arraybackslash}X} 

\begin{document}
%

\title{C-Rank: A Link-based Similarity Measure for Scientific Literature Databases}
%
%
%
%
%
\numberofauthors{3} 
%
\author{
%
%
\alignauthor Seok-Ho Yoon\\
       \affaddr{Dept. of Electronics and Computer Engineering}\\
       \affaddr{Hanyang University}\\
       \affaddr{Seoul, 133-791, Korea}\\
       \email{bogely@hanyang.ac.kr}
\alignauthor Sang-Wook Kim\\
       \affaddr{Dept. of Electronics and Computer Engineering}\\
       \affaddr{Hanyang University}\\
       \affaddr{Seoul, 133-791, Korea}\\
       \email{wook@hanyang.ac.kr}
\alignauthor Sunju Park\\
       \affaddr{School of Business, Yonsei University}\\
       \affaddr{Yonsei University}\\
       \affaddr{Seoul, 120-749, Korea}\\
       \email{boxenju@yonsei.ac.kr}
}

 \maketitle
\begin{abstract}

As the number of people who use scientific literature databases
grows, the demand for literature retrieval services has been
steadily increased. One of the most popular retrieval services is to
find a set of papers similar to the paper under consideration, which
requires a measure that computes similarities between papers.
Scientific literature databases exhibit two interesting
characteristics that are different from general databases. First,
the papers cited by old papers are often not included in the
database due to technical and economic reasons. Second, since a
paper references the papers published before it, few papers cite
recently-published papers. These two characteristics cause all
existing similarity measures to fail in at least one of the
following cases: (1) measuring the similarity between old, but
similar papers, (2) measuring the similarity between recent, but
similar papers, and (3) measuring the similarity between two similar
papers: one old, the other recent. In this paper, we propose a new
link-based similarity measure called C-Rank, which uses both in-link
and out-link by disregarding the direction of references. In
addition, we discuss the most suitable normalization method for
scientific literature databases and propose an evaluation method for
measuring the accuracy of similarity measures. We have used a
database with real-world papers from DBLP and their reference
information crawled from Libra for experiments and compared the
performance of C-Rank with those of existing similarity measures.
Experimental results show that C-Rank achieves a higher accuracy
than existing similarity measures.
\end{abstract}

\vspace{1mm} \noindent {\bf Categories and Subject Descriptors:}
I.5.3 {[Clustering]} {Similarity measures}

\vspace{1mm} \noindent {\bf General Terms:}{ Measurement,
Reliability}

\vspace{1mm} \noindent {\bf Keywords:}{ Scientific Literature,
Link-based Similarity Measure}

\section{Introduction}
As the number of people who use scientific literature databases
grows, the demand for scientific literature retrieval services has
been steadily increased. One of the most popular retrieval services
is to find a set of papers similar to the paper under consideration,
which requires a measure that computes similarities between papers.
Various similarity measures, either based on keywords or references,
have been proposed in the field of information retrieval
\cite{Bae99}. Text-based similarity measures count the number of
keywords in common between two papers. Link-based similarity
measures transform the reference information in a paper into
directed links and compute the similarity score between papers using
graph-based methods \cite{Kes63}\cite{Sma73}.

Intuitively, two scientific papers are considered similar when the
research problems dealt in those papers are similar. Text-based
similarity measures are not suitable in this regard, since they may
conclude two papers are similar as long as the context is similar
even when the problems the papers tackle are different \cite{Bae99}.
Link-based measures, on the other hand, use the reference created by
the authors to the papers that solve similar problems. Therefore,
similarity measures based on the reference information tend to be
more consistent with people's view on which papers are similar
\cite{Sha95}\cite{Zha09}. In this paper, we propose a new link-based
similarity measure for scientific literature databases.

There have been many link-based similarity measures in the
literature
\cite{Kes63}\cite{Sma73}\cite{Zha09}\cite{Ams72}\cite{Jeh02}\cite{Lib03}\cite{Lu01}\cite{Fog05}\cite{Yin06}\cite{Ant08}.
Typical link-based similarity measures include Bibliographic
Coupling (Coupling) \cite{Kes63}, Co-citation \cite{Sma73}, Amsler
\cite{Ams72}, rvs-SimRank \cite{Zha09}, SimRank \cite{Jeh02}, and
P-Rank \cite{Zha09}. In Co-citation, the similarity between two
objects is computed based on the number of objects that reference
both objects (i.e., in-link). The more objects that reference both
objects, the higher similarity score of two objects \cite{Sma73}. In
Coupling, the similarity between two objects is computed based on
the number of objects which are referenced by both of them (i.e.,
out-link). The more objects that are referenced by both objects, the
higher similarity score of two objects \cite{Kes63}. Amsler measures
the similarity between two objects as a weighted sum of the
similarity scores by Coupling and by Co-citation \cite{Ams72}.
SimRank improves the accuracy of Co-citation by computing the
similarity score iteratively. The iterative computation of
similarity captures the recursive intuition that two objects are
similar if they are referenced by similar objects \cite{Jeh02}.
Rvs-SimRank and P-Rank improves Coupling and Amsler, respectively,
in the similar way \cite{Zha09}.

Scientific literature databases exhibit two unique characteristics
that do not exist in general databases. First, few papers exist
which are referenced by old papers. This is because very old papers
are often not included in the database due to technical and economic
reasons. Second, since a paper can reference only the papers
published before it (and never the papers published after it), there
exist few papers which reference recently-published papers.

These two characteristics in a scientific literature database cause
all existing link-based similarity measures to fail in at least one
of the following three cases: (1) measuring the similarity between
old papers, (2) measuring the similarity between recent papers, and
(3) measuring the similarity between an old paper and a recent one.

First, Coupling, which uses out-link, may compute the similarity
score between two old but similar papers as near 0, because there
exist few papers that are referenced by both of them in the
database. Second, Co-citation, which uses in-link, on the other
hand, may compute the score between two recent but similar papers as
near 0, because there exist few papers which reference both papers
in the database. Third, both Coupling and Co-citation may compute
the score between two similar papers, one old and the other recent,
as near 0, because the old paper tends to have few papers that are
referenced by it and the recent one tends to have few papers that
reference it. Other similarity measures are plagued with similar
problems, which are discussed in detail in Section 2.

Two papers $p$ and $q$ should be determined similar in the following
three cases. First, $p$ and $q$ are similar if the number of papers
referenced by both $p$ and $q$ (out-links) is high. Second, $p$ and
$q$ are similar if the number of papers which reference both $p$ and
$q$ (in-links) is high. Third, $p$ and $q$ are similar if many of
the papers that are referenced by $p$ reference $q$. Though the
first and the second cases are captured in Coupling and Co-citation,
respectively, but they fail to address both cases simultaneously.
Moreover, no existing measures can be used for the third case.

To compute the similarity score correctly regardless of the
published dates of papers, one should consider all three cases
simultaneously. In other words, one should employ all three
measures: Coupling for computing the similarity between recent
papers, Co-citation for computing the similarity between old papers,
and a new measure for computing the similarity between an old and a
recent papers. This can be achieved by transforming both out-links
and in-links into undirected links and computing the similarity
based on the number of papers `connected' by two papers. In this
paper, we propose C-Rank, a new similarity measure that computes the
similarity properly for all three cases.

Existing similarity measures use various normalization methods to
prevent the similarity score between two papers from increasing as
the number of links to and from the papers increases
\cite{Jeh02}\cite{Fog05}\cite{Han06}. Typical normalization methods
include Jaccard coefficient, used in Coupling, Co-citation, and
Amsler, and the pairwise method, used in rvs-SimRank, SimRank, and
P-Rank. In this paper, we show that Jaccard coeffiecient is more
suitable than the pairwise method for scientific literature
databases through experiments.

The ideal similarity measure should match the intuition of users,
and the best way to evaluate similarity measures is to employ humans
\cite{Jeh02}. In this paper, we point out the problems with the
evaluation methods used in previous studies and propose a new method
that solves those problems. We use the proposed evaluation method in
our experiments.

The paper consists of the following. Section 2 points out the
problems with existing similarity measures when applied to
scientific literature databases. Section 3 describes C-Rank, the
detailed algorithm, and the suitable normalization method. Section 4
compares the accuracy of C-Rank with those of existing measures
through experiments. Section 5 summarizes and concludes the paper.

\section{Related Work}
In this section, we examine existing link-based similarity measures
and discuss why they fail to measure similarity correctly when used
for scientific literature databases.

\subsection{Link-Based Similarity Measures}

Existing link-based similarity measures include Co-citation,
Coupling, Amsler, SimRank, rvs-SimRank, and P-Rank \cite{Zha09}.
Co-citation, Coupling, and Amsler were proposed for measuring
similarity among scientific papers \cite{Zha09}, and were applied to
different types of objects with link information
\cite{Lar96}\cite{Pit97}\cite{Pop00}. SimRank, rvs-SimRank, and
P-Rank, on the other hand, were originally proposed for general
objects with link information \cite{Zha09}\cite{Jeh02}.

In Co-citation, the similarity between two objects is computed based
on the number of objects that have in-links to both objects.
Equation 1 represents Co-citation. $p$ and $q$ denote objects,
$S(p,q)$ the similarity score between $p$ and $q$, and $I(p)$ the
set of in-link neighbors of $p$.
\begin{equation}
S(p,q)=I(p) \cap I(q)
\end{equation}

In Coupling, the similarity between two objects is computed based on
the number of objects that have out-links from both objects.
Equation 2 represents Coupling. $O(p)$ denotes the set of out-link
neighbors of $p$. \vspace{-0.1cm}
\begin{equation}
S(p,q)=O(p) \cap O(q)
\end{equation}

Amsler measures the similarity between two objects as a weighted sum
of the similarity scores by Coupling and by Co-citation. Equation 3
represents Amsler. The relative weight of the similarity score of
Co-citation and that of Coupling is balanced by parameter $\lambda$.
In most applications, $\lambda$ is set at 0.5
\cite{Zha09}\cite{Ams72} . \vspace{-0.05cm}
\begin{equation}
S(p,q)= \lambda\times(I(p) \cap I(q))+(1-\lambda)\times(O(p) \cap
O(q))
\end{equation}

Figure 1 shows an example of a reference graph. $a$ to $j$ represent
papers and arrows represent reference relations between papers. The
similarity score between $e$ and $f$ by Co-citation is 1, because
there is one paper $i$ that references both papers. The score
between $e$ and $f$ by Coupling is 1, because a single paper $b$ is
referenced by both. The score between $e$ and $f$ by Amsler is 1,
assuming the relative weight for Coupling and Co-citation is 0.5.
\begin{figure}[h]
    \centerline{\psfig{figure=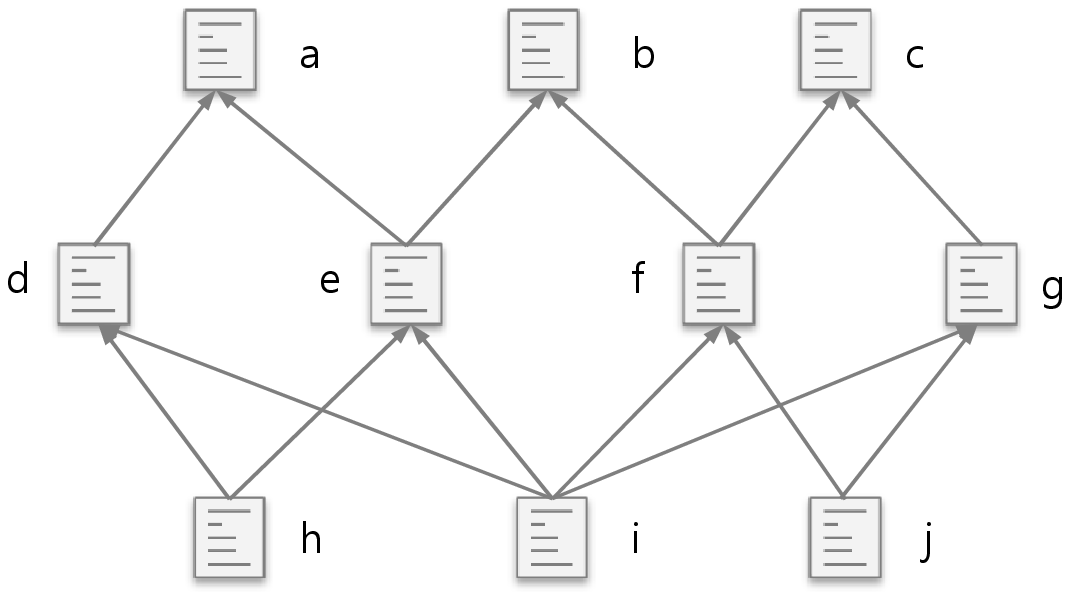,width=6cm} }
    \caption{A reference graph.}
    \label{fig:1}
\end{figure}

On the other hand, Co-citation computes the score between $a$ and
$c$ as 0 and the score between $d$ and $g$ as 1. A closer look
reveals that $d$ references $a$ and that $g$ references $c$. Since
the papers with the similarity score of 1 ($d$ and $g$) reference
them, $a$ and $c$ may be regarded somewhat similar. SimRank captures
this intuition such that the objects referenced by similar objects
are similar. That is, SimRank computes the similarity score
recursively. Equation 4 represents SimRank. In Equation 4,
$R_{k}(p,q)$ denotes the similarity score between $p$ and $q$ at
iteration $k$ , and $I_{i}(p)$ denotes the paper connected to $p$
through $i$-th in-link. $C$ is a decay factor for attenuating the
similarity score during similarity propagation, where $C \in [0,1]$.
\vspace{-0.1cm} {\begin{align}
&R_0(p,q)=\left\{\begin{array}{ll} 0 & \textrm{if $p \neq q$}\\1 & \textrm{if $p = q$}\end{array},\right.\nonumber\\[-1.5mm]
                    & \nonumber\\
&R_{k+1}(p,q)=\frac{C}{|I(p)||I(q)|}\sum_{i=1}^{|I(p)|}\sum_{j=1}^{|I(q)|}R_k(I_i(p),I_j(q))\nonumber\\[-1.5mm]
\label{eq:4}
\end{align}}

By using globalized neighbors, SimRank improves the accuracy of
Co-citation which uses localized neighbors only. Similarly,
rvs-SimRank and P-Rank improve Coupling and Amsler, respectively.
Equation 5 represents rvs-SimRank. The only difference between
rvs-SimRank and SimRank is the type of links used. Equation 6
represents P-Rank. As shown in Equation 6, P-Rank measures the
similarity score between two objects as a weighted sum of the
similarity scores by rvs-SimRank and SimRank.

{\begin{align}
&R_0(p,q)=\left\{\begin{array}{ll} 0 & \textrm{if $p \neq q$}\\1 & \textrm{if $p = q$}\end{array},\right.\nonumber\\[-1.5mm]
                    & \nonumber\\
&R_{k+1}(p,q)=\frac{C}{|O(p)||O(q)|}\sum_{i=1}^{|O(p)|}\sum_{j=1}^{|O(q)|}R_k(O_i(p),O_j(q))\nonumber\\[-1.5mm]
\label{eq:5}
\end{align}}
\vspace{-0.1cm} {\begin{align}
&R_0(p,q)=\left\{\begin{array}{ll} 0 & \textrm{if $p \neq q$}\\1 & \textrm{if $p = q$}\end{array},\right.\nonumber\\[-1.5mm]
                    & \nonumber\\
&R_{k+1}(p,q)=\lambda \times \frac{C}{|I(p)||I(q)|}\sum_{i=1}^{|I(p)|}\sum_{j=1}^{|I(q)|}R_k(I_i(p),I_j(q))\nonumber\\[-1.5mm]
                    & \nonumber\\
       & + (1-\lambda) \times \frac{C}{|O(p)||O(q)|}\sum_{i=1}^{|O(p)|}\sum_{j=1}^{|O(q)|}R_k(O_i(p),O_j(q))
\label{eq:6}
\end{align}}

Table 1 summarizes the existing similarity measures \cite{Zha09}.
When $k=1$, $C=1$, and $\lambda=1$ (or $\lambda=0$), Equation 6
represents Co-citation (or Coupling). When $k=1$, $C=1$ and
$\lambda=0.5$, Equation 6 represents Amsler. When $k = \infty$,
Equation 6 represents SimRank, rvs-SimRank, and P-Rank depending on
the value of $\lambda$. Even though $k=\infty$ for SimRank,
rvs-SimRank, and P-Rank, empirically the similarity scores by
SimRank, rvs-SimRank, and P-Rank tend to converge at $k = 4$ or $5$
\cite{Zha09}\cite{Jeh02}.
\begin{table}[h]
    \centering
    \caption{Relationship among similarity measures (Adopted from \cite{Zha09})}
    \begin{tabular}{|c|c|c|c|}
        \hline
        Links used k
&In-link&Out-link&Both\\
        \hline
        \multirow{2}*[-.3ex]{k=1}&Co-citation&Coupling&Amsler \\
        &C=1, $\lambda$=1&C=1, $\lambda$=0&C=1, $\lambda$=1/2\\
        \hline
        \multirow{2}*[-.3ex]{k=$\infty$}&SimRank&rvs-SimRank&P-Rank \\
        &C=varies, $\lambda$=1&C=varies, $\lambda$=0&C, $\lambda$=varies \\
        \hline
    \end{tabular}
\end{table}

\subsection{Problems with Existing Similarity Measures}
Scientific literature databases have two characteristics that are
different from general databases. First, very old papers are often
not in the database. Second, there exist few papers that reference
recently-published papers. Due to these two characteristics, all
existing similarity measures fail to compute the similarity score
correctly in scientific literature databases, at least in one of the
following three cases.

\vspace{0.3cm}\begin{tabular}{|>{\centering}p{0.5cm}p{6.9cm}|}
    \hline
    (P1)&measuring the similarity between old, but similar papers\\
    (P2)&measuring the similarity between recent, but similar papers\\
    (P3)&measuring the similarity between two similar papers: one old, the other recent\\
    \hline
\end{tabular}
\vspace{0.3cm}

Figure 2 represents the reference relations among papers as a graph.
In Figure 2, $a$ to $l$ represent papers, and arrows represent the
reference relations between papers. The papers on top of the figure
are older, and the papers at bottom are more recent. An example of
(P1) happens when the similarity score between $a$ and $b$ is
computed. The similarity score computed by Coupling (rvs-SimRank) is
0 (near 0) because these papers have no out-links. The similarity
score by Amsler (P-Rank) is not 0, because the score by Co-citation
is 1. The maximum score by Amsler (P-Rank), however, would be at
most 0.5 (assuming the relative weight for Coupling and Co-citation
is 0.5). That is, the score by Amsler (P-Rank) is inaccurate. An
example of (P2) happens when the score between $k$ and $l$ is
computed. The score computed by Co-citation (SimRank) is 0 (near 0)
because these papers have no in-links. The score by Amsler (P-Rank)
would be 0.5 (near 0.5), even though they have a common out-link
neighbor $i$. An example of (P3) happens when the score between $e$
and $l$ is computed. The score computed by all existing similarity
measures is 0 or near 0.

\begin{figure}[h]
    \centerline{\psfig{figure=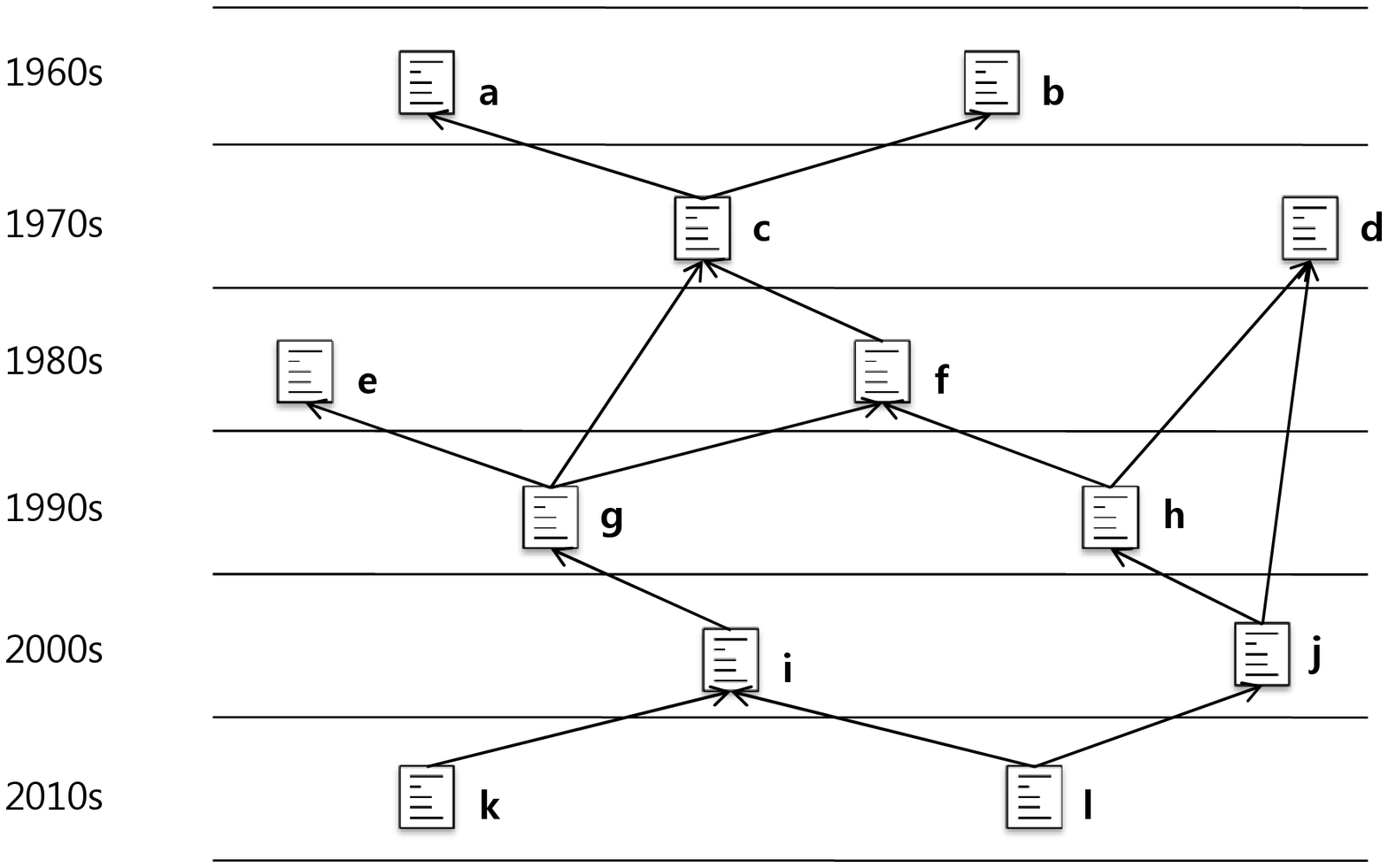,width=7cm} }
    \caption{A graph of the reference relationships with publishing dates.}
    \label{fig:2}
\end{figure}

Coupling, Co-citation, and Amsler fail to capture the similarity
between papers in scientific literature databases. Rvs-SimRank,
SimRank, and P-Rank are plagued with the same problems, since they
are the iterative extensions of Coupling, Co-citation, and Amsler,
respectively.

\section{Proposed Similarity Measure}

In this section, we propose a new similarity measure called C-Rank
and describe its algorithm in detail. We also discuss a
normalization method appropriate for the new measure.

\subsection{Main Idea}

Two papers $p$ and $q$ should receive a high similarity score in the
following three cases.

\begin{tabular}{|>{\centering}p{0.5cm}p{6.9cm}|}
    \hline
    (C1)&the number of papers referenced by both $p$ and $q$ is high\\
    (C2)&the number of papers which reference both $p$ and $q$ is high\\
    (C3)&the number of the papers which are referenced by $p$ reference $q$ is high\\
    \hline
\end{tabular}
\vspace{0.2cm}

We define the paper which is referenced by both papers as OP (common
Out-link Paper), paper which references both papers as IP (common
In-link Paper), and paper which is referenced by the one paper and
references the other as BP (common Between Paper). In Figure 2, for
example, $f$ is an OP of $g$ and $h$, $h$ is an IP of $d$ and $f$,
and $c$ is a BP of $a$ and $f$.

The existing measures can be used in (C1) and (C2) cases.
Co-citation or SimRank can be used for (C1), and Coupling or
rvs-SimRank can be used for (C2). In Figure 2, for example,
Co-citation (SimRank) can be used to measure the similarity between
$g$ and $h$, and Coupling (rvs-SimRank) can be used to measure the
similarity between d and $f$. The existing measures, however, cannot
correctly measure the similarity in (C3). In Figure 2, for example,
existing measures fail to compute the similarity between $a$ and
$f$. A similarity measure that counts BPs should be suitable for
this case. Of course, a BP-based similarity measure cannot be used
for the papers with publication dates close to each other, such as
$g$ and $h$, since there exist few BPs between the papers under
consideration. \vspace{0.2cm}

To compute the score correctly in all three cases, therefore, we
propose to use all three measures, Co-citation (or SimRank),
Coupling (or rvs-SimRank), and a new BP-based measure. When
combining all three measures, a weighted sum of similarity scores
from the three measures could have been used, similar to Amsler (or
P-Rank). Note that this would suffer the same problem faced by
Amsler (or P-Rank) that one of the scores may be near 0, which
results in the score that is much lower than the correct value.
Instead of using a weighted sum, therefore, we propose a new measure
that considers three cases simultaneously.

\subsection{C-Rank}
Though papers are classified into OPs, IPs, and BPs based on the
direction of links, their role is the same: they are used to compute
the similarity between two papers. So, we disregard the direction of
links, which results in a single type of links that connect two
papers. We define the papers which connect two papers as
$Connectors$. When disregarding the direction of references,
Coupling (or rvs-SimRank), Co-citation (or SimRank), and a BP-based
similarity measure are unified as a single measure that computes the
similarity score based on the number of Connectors in an undirected
graph.

We propose a Connector-based similarity measure called C-Rank.
C-Rank uses both in-links and out-links at the same time. Equation 7
represents C-Rank, where $L(p)$ denotes the set of undirected link
neighbors of paper $p$. Similar to that the accuracy of Co-citation
(Coupling) is improved by iterative SimRank (rvs-SimRank), C-Rank is
defined iteratively.

{\begin{align}
&R_0(p,q)=\left\{\begin{array}{ll} 0 & \textrm{if $p \neq q$}\\1 & \textrm{if $p = q$}\end{array},\right.\nonumber\\[-1.5mm]
& \nonumber\\
&R_{k+1}(p,q)=\frac{C}{|L(p)||L(q)|}\sum_{i=1}^{|L(p)|}\sum_{j=1}^{|L(q)|}R_k(L_i(p),L_j(q))
\label{eq:7}
\end{align}}

Unlike Amsler or P-Rank, C-Rank does not need parameter $\lambda$,
because C-Rank unifies in-links and out-links into undirected links.
Furthermore, C-Rank has the effect similar to increasing the weight
of Co-citation (SimRank) when computing the score between old
papers, increasing the weight of Coupling (rvs-SimRank) when
computing the score between recent papers, and increasing the weight
of a BP-based similarity measure when computing the score between
old and recent papers. The user does not have to set the value of
$\lambda$ when using C-Rank. In experiments, we show that the
accuracy of C-Rank is higher than those of Amsler (P-Rank) with
different $\lambda$ values.

One of the evaluation criteria for link-based similarity measures is
how many pairs of objects can be measured \cite{Zha09}. SimRank
fails to compute the similarity when a paper has no in-link, and
rvs-SimRank fails when a paper has no out-link. Although being able
to compute the similarity scores for more pairs than any other
measures, P-Rank measures similarity for less number of pairs than
C-Rank. This is because P-Rank fails to compute the similarity
between an old paper and a recent one. In experiments, we show that
the number of pairs of papers computed by C-Rank is more than that
of any other measures.

Treating both in-links and out-links as undirected might be thought
to result in loss of semantics of the direction of links. By
disregarding the direction of links, however, C-Rank is able to
consider all three cases mentioned in 2.2. Thus, the measure has
more advantages than disadvantages when computing the similarities
among papers.

\subsection{Normalization}
In previous studies, two types of normalization methods are used to
prevent a problem that the similarity score between two papers
increases as the number of links increases. Used in Coupling (or
Co-citation), Jaccard coefficient normalizes the similarity score by
dividing the number of papers which are referenced by (or reference)
both papers by the sum of the number of the papers each paper
references (or is referenced by) \cite{Han06}. Rvs-SimRank, SimRank,
and P-Rank have used the pairwise normalization method. SimRank, for
example, builds a set of pairs between the papers that reference any
one of the two papers under consideration, computes the sum of
similarity scores of all pairs, and divides it by the product of the
number of in-links to each paper.

In scientific literature databases, some well-known papers are
referenced by the many other papers, and people who use retrieval
services would be interested in those quality papers. Since the
pairwise normalization method lowers the similarity score of the
papers with many in-links, the similarity scores between two famous
papers can be very low \cite{Fog05}. Figure 3 represents an example
of the problem with the pairwise normalization method. In Figure 3,
papers $p$ and $q$ are referenced by all the other papers, and
should be determined similar. When the number of papers which
reference both $p$ and $q$ is $k$, however, the similarity score
with pairwise normalization becomes $\frac{1}{k}$. The same problem
exists when the similarity is computed iteratively, although the
score may be somewhat higher than $\frac{1}{k}$ \cite{Fog05}. So,
for the scientific literature databases where famous papers (in
which users would be interested) have many in-links, Jaccard
coefficient seems a better normalization method. Equation 8
represents C-Rank with Jaccard coefficient. In Equation 8,
`$\setminus$' denotes different set. In experiments, we show Jaccard
coefficient is more suitable than pairwise normalization for
scientific literature databases.

\begin{figure}[h]
    \centerline{\psfig{figure=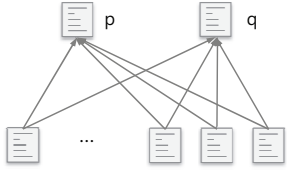,width=4cm} }
    \caption{An example showing the problem with the pairwise normalization method.}
    \label{fig:3}
\end{figure}

\subsection{Recursive C-Rank}

The recursive C-Rank in Equation (8) has the following four
properties. For any papers $p$ and $q$, the iterative C-Rank of $p$
and $q$ is the same as that of $q$ and $p$ (symmetry). The iterative
C-Rank is non-decreasing during similarity computation
(monotonicity). Existence and uniqueness guarantee that there exists
a unique solution to iterative C-Rank which reaches a fixed point by
iterative computation. The prove can be found in Appendix.

\vspace{0.2cm}
\begin{tabular}{|p{2cm}p{5.4cm}|}
\hline
(Symmetry)&$R_{k}(p,q) = R_{k}(q,p)$ \\
(Monotonicity)&$0 \leq R_{k}(p,q) \leq R_{k+1}(p,q) \leq 1$ \\
(Existence)&The solution to the iterative C-Rank equations always
exists and converges to a fixed point, $s(*,*)$, which is the
theoretical solution to the recursive C-Rank equations. \\
(Uniqueness)&The solution to the iterative C-Rank equation is unique
when $C\neq1$.\\\hline
\end{tabular}

\begin{align}
R_0(p,q)& =\left\{\begin{array}{ll} 0 & \textrm{if $p \neq q$}\\1 & \textrm{if $p = q$}\end{array},\right.\nonumber\\[-1.5mm]
                    & \nonumber\\
R_{k+1}(p,q) & = C \times (\frac{|L(p) \cap L(q)|}{|L(p) \cup L(q)|} \times 1\nonumber\\[-1.5mm]
                    & \nonumber\\
                    & + \frac{1}{|L(p) \cup L(q)||L(q)|} \times \sum_{p' \in L(p) \setminus L(q)}\sum_{q' \in L(q)}R_{k}(p',q') \nonumber&&&&&&&\\[-1.5mm]
                    & \nonumber\\
                    & \nonumber\\
                    & +  \frac{1}{|L(p) \cup L(q)||L(p)|} \times \sum_{p' \in L(p)}\sum_{q' \in L(q) \setminus L(p)}R_{k}(p',q') \nonumber\\[-1.5mm]
                    & \nonumber\\
\label{eq:8}
\end{align}

\subsection{Algorithm}

Table 2 shows the algorithm of C-Rank. For every pair of papers
$(p,q)$, an entry $R(p,q)$ maintains the intermediate C-Rank score
of (a,b) during iterative computation. Because the $k$-th iterative
C-Rank score is computed based on C-Rank scores in the $(k-1)-$th
iteration, an auxiliary similarity score store $R*(p,q)$ is
maintained accordingly. The code first initializes $R_{0}(p,q)$
based on Table 2 (Lines 1$\sim$4). During iterative computation,
$R*(*,*)$, is updated by $R(*,*)$ in the $k-1$ iteration, based on
Table 2 (Lines 6$\sim$17). Then $R_{k}(*,*)$ is substituted by
$R_{k+1}(*,*)$ for further iteration (Lines 18$\sim$20). This
iterative procedure is repeated $k$ times (Lines 5$\sim$21).

The space complexity of all existing measures are $O(n^{2})$ because
the measures must store pairs of all papers. Let $d^{1}$ and $d^{2}$
be the average number of in-links and out-links of all papers,
respectively, the time complexity of rvs-SimRank, SimRank, and
P-Rank are $O(k\cdot d_{1}^{2}\cdot n^{2})$, $O(k\cdot
d_{2}^{2}\cdot n^{2})$, and $O(k\cdot(d_{1}^{2}+d_{2}^{2})\cdot
n^{2})$), respectively \cite{Zha09}. The time complexity of C-Rank
is $O(k\cdot(d_{1}+d_{2})^2\cdot n^{2})$, which is slightly higher
than the others. However, the worst case time complexity of all
existing iterative measures including C-Rank is $O(n^{4})$.

The time complexity of C-Rank may become too high. There have been
many methods to improve the time complexity of SimRank
\cite{Zha09}\cite{Jeh02}\cite{Fog05}\cite{Yin06}\cite{Ant08}\cite{Liz08}.
These methods can be applied to C-Rank, because the equation of
C-Rank and that of SimRank are similar.


\section{Experiments}

In this section, we compare the effectiveness of C-Rank and the
existing similarity measures.
\subsection{Experimental Setup}
Our experiments ran on a scientific literature database with papers
from DBLP\footnote{http://www.informatic.uni-trier.de/$\sim$ley/db/}
and reference information crawled from
Libra\footnote{http://academic.research.microsoft.com}. We used the
papers related to the database research because the running time of
the existing similarity measures and C-Rank in the large database
can become very high. We used the publication venues listed in
\cite{Yan07} to select papers related to database research. Table 3
lists the publication venues in \cite{Yan07}. The number of papers
was 23,795 and the number of references (to the papers in the
dataset) was 126,281. All our experiments were performed on an Intel
PC with Quad Core 2.67GHz CPU, running Windows 2008. We compared
C-Rank with rvs-SimRank, SimRank, and P-Rank, because Coupling,
Co-citation, and Amsler could be expressed using rvs-SimRank,
SimRank, and P-Rank, respectively. For fairness of comparison, we
set the decay factor $C=0.8$ for all measures and the relative
weight $\lambda$ to be 0.5 for P-Rank, unless otherwise noted. All
the default values of parameters are set in accordance with
\cite{Zha09}.

\begin{table}[h]
\centering \caption{The C-Rank Algorithm}\label{tab-2}
\begin{tabular}{|rl|}
\hline
        & C-Rank ($G$, $C$, $k$) \\\hline
        & Input: A reference graph $G$ (an undirected graph), \\
        & the decay factor $C$, the iteration number $k$\\
        & Output: C-Rank score $R(*,*)$\\\hline
        1 & foreach $p \in G$ do /* Initialization */ \\
                \vspace{-0.35cm}
        & \\
        \vspace{-0.35cm}
        2 &\hspace{0.2cm} foreach $q \in G$ do\\
        & \\
        \vspace{-0.35cm}
        3      &\hspace{0.4cm}  if $p==q$ then $R(p,q)=1$\\
        & \\
        \vspace{-0.35cm}
        4       &\hspace{0.4cm} else $R(p,q)=0$\\
        & \\
        \vspace{-0.35cm}
        5 & while ($n<k$) do  /* Iteration */\\
        & \\
        \vspace{-0.35cm}
        6   &\hspace{0.2cm} foreach $p \in G$ do\\
        & \\
        \vspace{-0.35cm}
        7      &\hspace{0.4cm}  foreach $q \in G$ do\\
        & \\
        \vspace{-0.35cm}
        8           &\hspace{0.6cm}  $R*(p,q) = \frac{|L(p) \cap L(q)|}{|L(p) \cup L(q)|}$ \\
        & \\
        \vspace{-0.35cm}
        9           &\hspace{0.6cm} foreach $l_{p} \in L(p)\backslash L(q)$\\
        & \\
        \vspace{-0.35cm}
        10              &\hspace{0.8cm} foreach $l_{q} \in L(q)$\\
        & \\
        \vspace{-0.35cm}
        11                  &\hspace{1cm}  differentSetofp  += $R(l_{p},l_{q})$\\
        & \\
        \vspace{-0.35cm}
        12              &\hspace{0.8cm} differentSetofp $\times$= $\frac{1}{|L(p) \cup L(q)||L(q)|}$\\
        & \\
        \vspace{-0.35cm}
        13             &\hspace{0.6cm}  foreach $l_{p} \in L(q)\backslash L(p)$\\
        & \\
        \vspace{-0.35cm}
        14                  &\hspace{0.8cm} foreach $l_{q} \in L(p)$\\
        & \\
        \vspace{-0.35cm}
        15                      &\hspace{1cm}  differentSetofq  +=  $R(l_{p},l_{q})$\\
        & \\
        \vspace{-0.35cm}
        16              &\hspace{0.8cm} differentSetofq $\times$= $\frac{1}{|L(p) \cup L(q)||L(p)|}$ \\
        & \\
        \vspace{-0.35cm}
        17         &\hspace{0.6cm} $R*(p,q) += C \times$ (differentSetofp + differentSetofq)\\
        & \\
        \vspace{-0.35cm}
        18  & \hspace{0.2cm}foreach $p \in G$ do /* Update */ \\
        & \\
        \vspace{-0.3cm}
        19      &\hspace{0.4cm} foreach $q \in G$ do\\
        & \\
        \vspace{-0.35cm}
        20 &\hspace{0.6cm} $R(p,q) = R*(p,q)$\\
        & \\
        \vspace{-0.35cm}
        21 & $n=n+1$\\
        & \\
        \vspace{-0.35cm}
        22 & return R(*,*)\\
        & \\
\hline
\end{tabular}
\end{table}

\subsection{Accurate Evaluation Method}
Previous studies on similarity measures used various evaluation
methods. \cite{Kes63} and \cite{Sma73} evaluated Coupling and
Co-citation qualitatively, showing some example cases. Although easy
to use, however, qualitative evaluations do not provide any concrete
evidence on which measure is better or how accurate each measure is.
\cite{Jeh02} used a text-based similarity measure and Co-citation as
ground truth to evaluate the accuracy of SimRank. Because the
text-based similarity measure is less accurate than SimRank, and
Co-citation does not generate similarity scores accurately at least
in scientific literature databases, using these two measures as
ground truth do not seem a good evaluation method for scientific
literature databases. \cite{Zha09} clustered papers using the
similarity score by SimRank and the similarity score by P-Rank,
respectively, and evaluated the accuracy of two measures by
comparing the similarity scores of papers from the same cluster and
those from different clusters. Although used for evaluating the
quality of clustering in clustering research, this method is not
suitable for evaluating the similarity measure because the results
are dependent on the type of data and clustering algorithm
\cite{Xu05}.

\begin{table}[h]
\centering \caption{Publication venues related to database research
\cite{Yan07}}\label{tab-3}
\begin{tabular}{|p{8cm}|}
\hline ADBIS, ADC, ARTDB, BNCOD, CDB, CIKM, CoopIS, DANTE, DASFAA,
DAWAK, DB, DBPL, DBSEC, DEXA, DKD, DKE, DL, DMKD, DNIS, DOLAP, DOOD,
DPD, DPDS, DS, EDBT, ER, FODO, FOIKS, FQAS, GIS, HPTS, ICDE, ICDM,
ICDT, ICIS, IDA, IDEAL, IDEAS, IGSI, Inf. Process, Lett., Inf. Sci.,
Inf. Syst., IPM, IQIS, ISF, ISR, IW-MMDBMS, IWDM, JDM, JIIS, JMIS,
K-CAPKA, KDD, KER, KIS, KR, MDA, MFDBS, MLDM, MMDB, MSS, NLDB,
OODBS, PAKDD, PKDD, PODS, RIDE, RIDS, SIGKDD Exp., SIGMOD, SIGMOD
Rec., SSD, SSDMB, TKDE, TODS, TOIS, TSDM, UIDIS, VDB, VLDB, VLDB-J,
WebDB, WIDM, WISE, XMLEC\\\hline
\end{tabular}
\end{table}

One of the most accurate ways to evaluate the accuracy of a
similarity measure would be to ask humans \cite{Jeh02}, but user
studies are expensive and time consuming. We propose a new
evaluation method that achieves similar effects without employing
user studies. We ask domain experts to select the papers similar to
each other, and evaluate each similarity measure based on the
similarity score between the selected papers. The higher the score
is, the more accurate the similarity measure is.

The evaluation process in detail is as follows. First, we select
five well-known fields in data mining (clustering, sequential
pattern mining, graph mining, spatial databases, link mining) and
select the references at the end of each chapter for each field from
a data mining text book \cite{Han06}. The references include both
old and recent papers. Second, we use one of the references to be a
query paper and find the $m$ highest scoring papers (where $m$ can
be 10, 20, 30, 40, and 50) by each similarity measure. Third, we
compute the precision of each similarity measure by comparing the m
highest scoring papers to those in the reference list of the field
of the query paper. Fourth, we repeat the second and third steps
until all references are used as a query paper.

\subsection{Experimental Results}
\subsubsection{Normalization Method}
In this section, we compare the accuracy of similarity measures with
Jaccard coefficient and that with the pairwise normalization method.
Figure 4 shows the accuracy of P-Rank and C-Rank with different
normalization methods. (The other measures, rvs-SimRank and SimRank,
exhibit similar results, and thus omitted.) The accuracy of both
similarity measures with Jaccard coefficient is higher than that
with pairwise normalization. The results confirm that Jaccard
coefficient is a more suitable normalization method for scientific
literature databases. Note that the accuracy of C-Rank with pairwise
normalization is lower than that of P-Rank with pairwise
normalization. This is because C-Rank uses more links than P-Rank as
mentioned 3.3.


\subsubsection{Top 10 Rankings}
In this section, we confirm that C-Rank measures similarity properly
by extracting the top 10 highest-scoring papers by C-Rank when
paired with a well-known paper as a query paper. We use \cite{Gut84}
and \cite{Zha96}, two well-known papers in the database and data
mining research field, respectively. Table 4 lists top 10
highest-scoring papers when paired with \cite{Gut84}, and Table 5
lists the top 10 highest-scoring papers when paired with
\cite{Zha96}. \cite{Gut84} proposed R-Tree as a multidimensional
index. In Table 4, the highest-scoring papers by C-Rank are mostly
related to multidimensional indexes. \cite{Zha96} proposed BIRCH as
a clustering method. In Table 5, the papers by C-Rank are mostly
related to clustering. The results show that C-Rank can provide a
set of papers similar to the paper under consideration.
\vspace{0.2cm}
\begin{figure}[h]
    \centerline{\psfig{figure=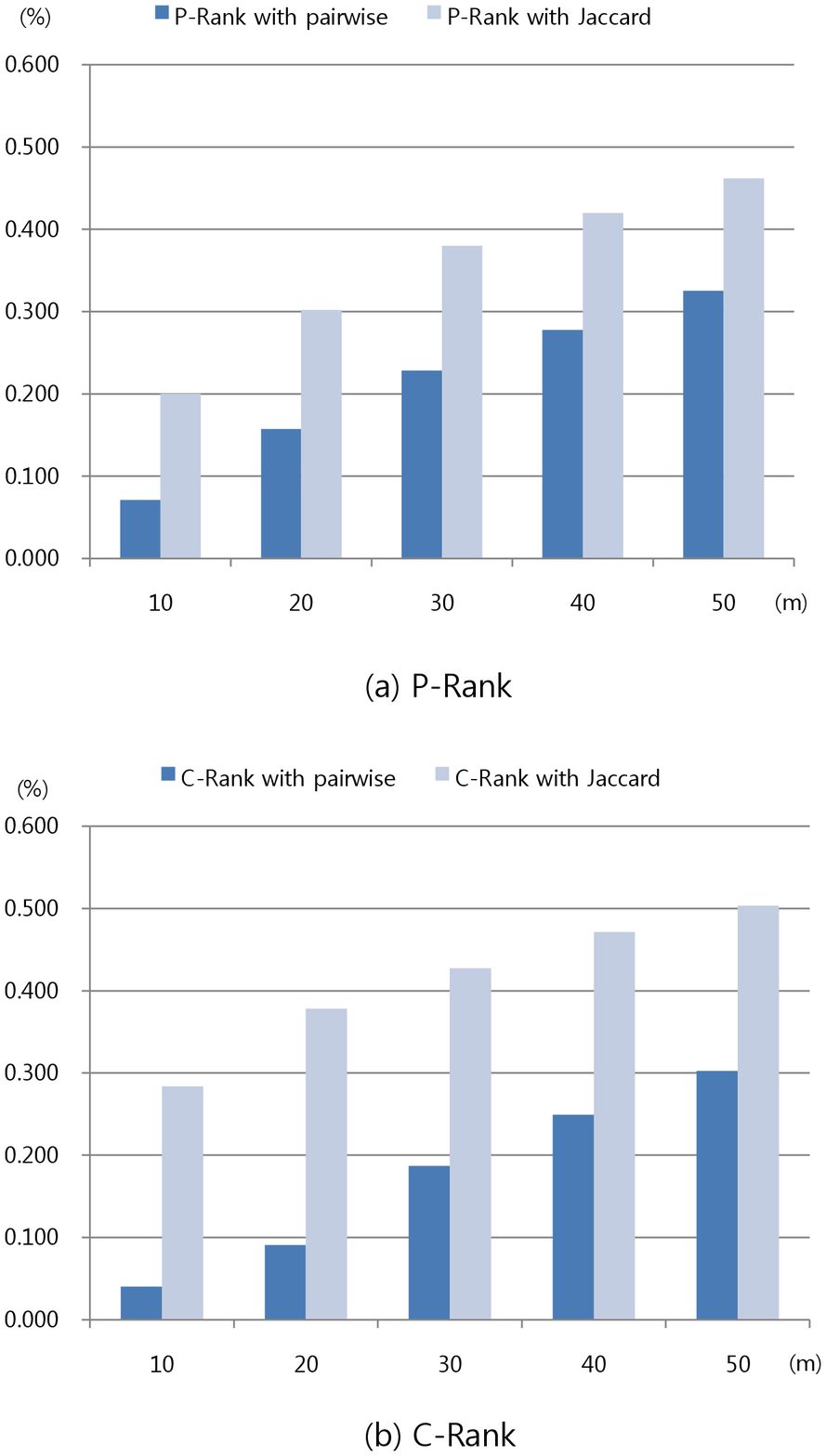,width=6cm} }
    \caption{Comparing Jaccard coefficient and pairwise normalization method.}
    \label{fig:4}
\end{figure}

\subsubsection{Failure of Existing Similarity Measures}
In this section, we demonstrate the problem of existing similarity
measures when applied to scientific literature databases using three
cases identified in Section 2.2. We also show that C-Rank computes
the similarity score properly in all three cases. For demonstration
purposes, we select {\cite{Kno71}, \cite{Gho75}} and {\cite{Gut84},
\cite{Rob81}} as the pairs of old papers but similar papers,
{\cite{Wan04}, \cite{Nij04}} and {\cite{Con05}, \cite{Che04}} as the
pairs of recent papers but similar papers, and {\cite{Rob81},
\cite{Atl04}} and {\cite{Ng05}, \cite{Rus73}} as the pairs of an old
and a recent paper.

Table 6 shows the result of case analysis. Six cases are illustrated
in Table 6, but all other examples tested show similar results. In
Table 6, the similarity scores between old but similar papers by
rvs-SimRank are 0 in both cases. As noted in Section II.B,
rvs-SimRank identifies incorrectly that the papers are not similar
because they have no common out-links. Similarly, the similarity
scores between recent but similar papers by SimRank are 0 in both
cases. SimRank identifies incorrectly that the papers are not
similar because they have no common in-links. Furthermore, all
existing similarity measures compute the similarity scores between
the papers with different publication dates as 0. C-Rank is the only
one that measures the similarity of those papers. That is, C-Rank is
able to capture the similarity between the papers with different
publishing dates. Note that the scores by C-Rank are not high in
both cases. This is because the problem tackled in the old paper and
that in the newer paper, although somewhat similar, have become less
in common as time passes on. The original problem may have changed
to a more specific problem, or it may have changed to solve more
general problem, etc.
\begin{table}[h]
\centering \caption{Top 10 papers similar to \cite{Gut84}}
\begin{tabular}{|c|l|} \hline
{1}& The R*-Tree: An Efficient and Robust Access Method ...\\
\hline
{2}& The R+-Tree: A Dynamic Index for Multi-Dimensional ...\\
\hline
{3}& Nearest Neighbor Queries\\
\hline
{4}& The K-D-B-Tree: A Search Structure For Large ...\\
\hline
{5}& The X-tree : An Index Structure or ...\\
\hline
{6}& On Packing R-trees\\
\hline
{7}& The Grid File: An Adaptable, Symmetric Multikey ...\\
\hline
{8}& Efficient Processing of Spatial Joins Using R-Trees\\
\hline
{9}& Hilbert R-tree: An Improved R-tree using Fractals\\
\hline
{10}& The SR-tree: An Index Structure for High-Dimensional ...\\
\hline
\end{tabular}
\end{table}

\begin{table}[h]
\centering \caption{Top 10 papers similar to \cite{Zha96}}
\begin{tabular}{|c|l|} \hline
{1}& Efficient and Effective Clustering Methods ...\\
\hline
{2}& CURE: An Efficient Clustering Algorithm ...\\
\hline
{3}& A Density-Based Algorithm for Discovering Clusters ...\\
\hline
{4}& Automatic Subspace Clustering of High Dimensional ...\\
\hline
{5}& Scaling Clustering Algorithms to Large Databases\\
\hline
{6}& WaveCluster: A Multi-Resolution Clustering Approach ...\\
\hline
{7}& Fast Algorithms for Projected Clustering\\
\hline
{8}& STING: A Statistical Information Grid Approach ...\\
\hline
{9}& An Efficient Approach to Clustering in Large ...\\
\hline
{10}& OPTICS: Ordering Points To Identify the Clustering...\\
\hline
\end{tabular}
\end{table}

\begin{table}[h]
    \centering
    \caption{The results of case analysis}
    \begin{tabular}{|c|c|c|c|}
    \hline
    & \multirow{2}{*}{old papers} & \multirow{2}{*}{recent papers} & an old\\
    & & & and a recent paper\\
    \hline
    &  \cite{Kno71} and \cite{Gho75} &  \cite{Wan04} and \cite{Nij04} &  \cite{Rob81} and \cite{Atl04}\\
    &  \cite{Gut84} and \cite{Rob81} &  \cite{Con05} and \cite{Che04} &  \cite{Ng05} and \cite{Rus73}\\
    \hline
     \multirow{2}{*}{rvs-SimRank} & 0 & 0.278 & 0\\
    & 0 & 0.189 & 0\\
    \hline
     \multirow{2}{*}{SimRank} & 0.179 & 0 & 0\\
    & 0.141 & 0 & 0\\
    \hline
     \multirow{2}{*}{P-Rank} & 0.114 & 0.198 & 0\\
    & 0.082 & 0.096 & 0\\
    \hline
     \multirow{2}{*}{C-Rank} & 0.240 & 0.282 & 0.050\\
    &0.175 & 0.210 & 0.047\\
    \hline
    \end{tabular}
\end{table}

\subsubsection{Accuracy of Similarity Measures}
Figure 5 represents the accuracy of different similarity measures.
In Figure 5, x-axis represents the number of top $m$ scoring papers,
and y-axis represents the accuracy of each similarity measure. As
shown in Figure 5, the accuracy of C-Rank is higher than the other
similarity measures regardless of the value of $m$. The results
indicate that C-Rank is more accurate than the other measures in
scientific literature databases. \vspace{0.2cm}

\subsubsection{Distribution of Similarity Scores}
In this section, we count the number of pairs whose similarity is
computable by each similarity measure. Figure 6 shows the
distribution of the similarity scores by each similarity measure. In
Figure 6, x-axis represents the range of similarity scores, where
[$lb$, $ub$) indicates $lb$ is included and $ub$ is not included in
the range, and y-axis represents the number of pairs of papers. In
Figure 6, y-axis is in \emph{log scale}, because for most pairs, the
similarity scores are either in \emph{N/A} or in [0, 0.1).
\emph{N/A} represents the pairs whose similarity cannot be measured.
As shown in Figure 6, there are no such pairs of papers whose
similarity scores are \emph{N/A} by C-Rank. This implies that C-Rank
computes the similarity score between all pairs of papers because
C-Rank uses both in-link and out-link simultaneously. In Figure 6,
the pairs of papers whose similarity scores are \emph{N/A} by the
other measures can be thought to be computed as near 0 by C-Rank.
However, we note that the number of pairs in [0, 0.1) by C-Rank is
not too much different from those of other measures. This result
indicates that C-Rank provides meaningful similarity scores for the
pairs of papers even when their computation is infeasible with the
other similarity measures. \vspace{0.2cm}

\begin{figure}[h]
    \centerline{\psfig{figure=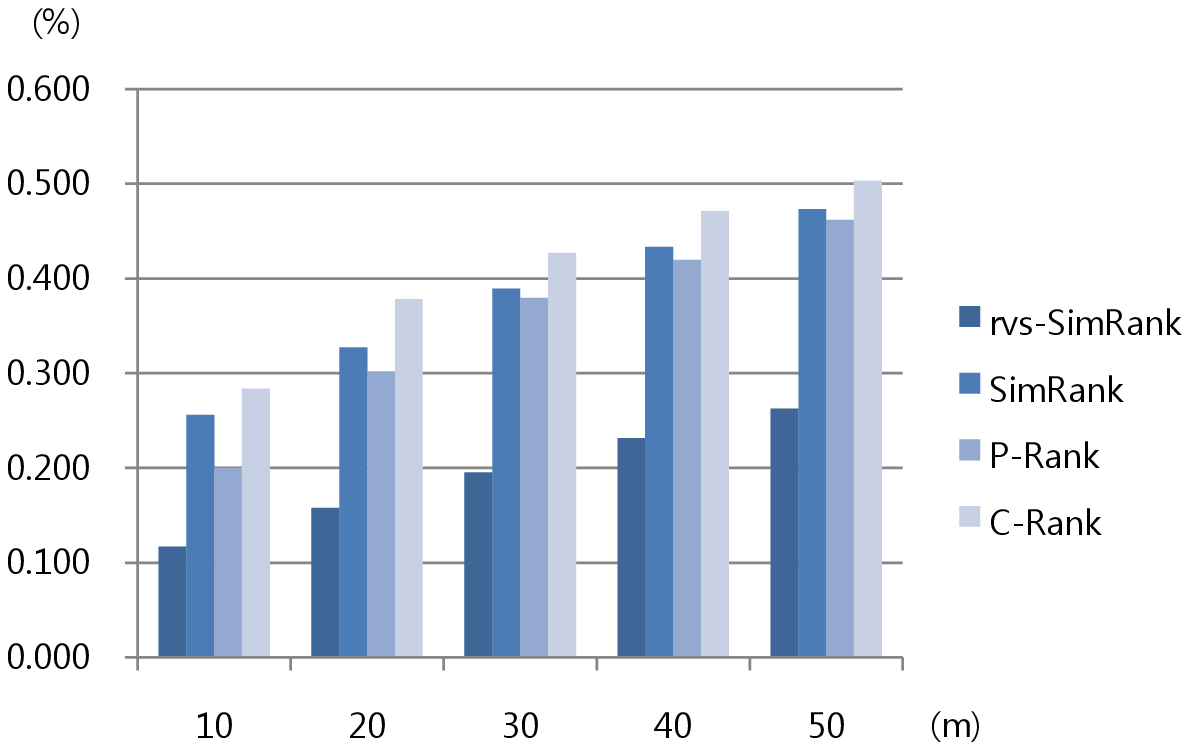,width=8cm} }
    \caption{The accuracy of the similarity measures.}
    \label{fig:5}
\end{figure}

\begin{figure}[h]
    \centerline{\psfig{figure=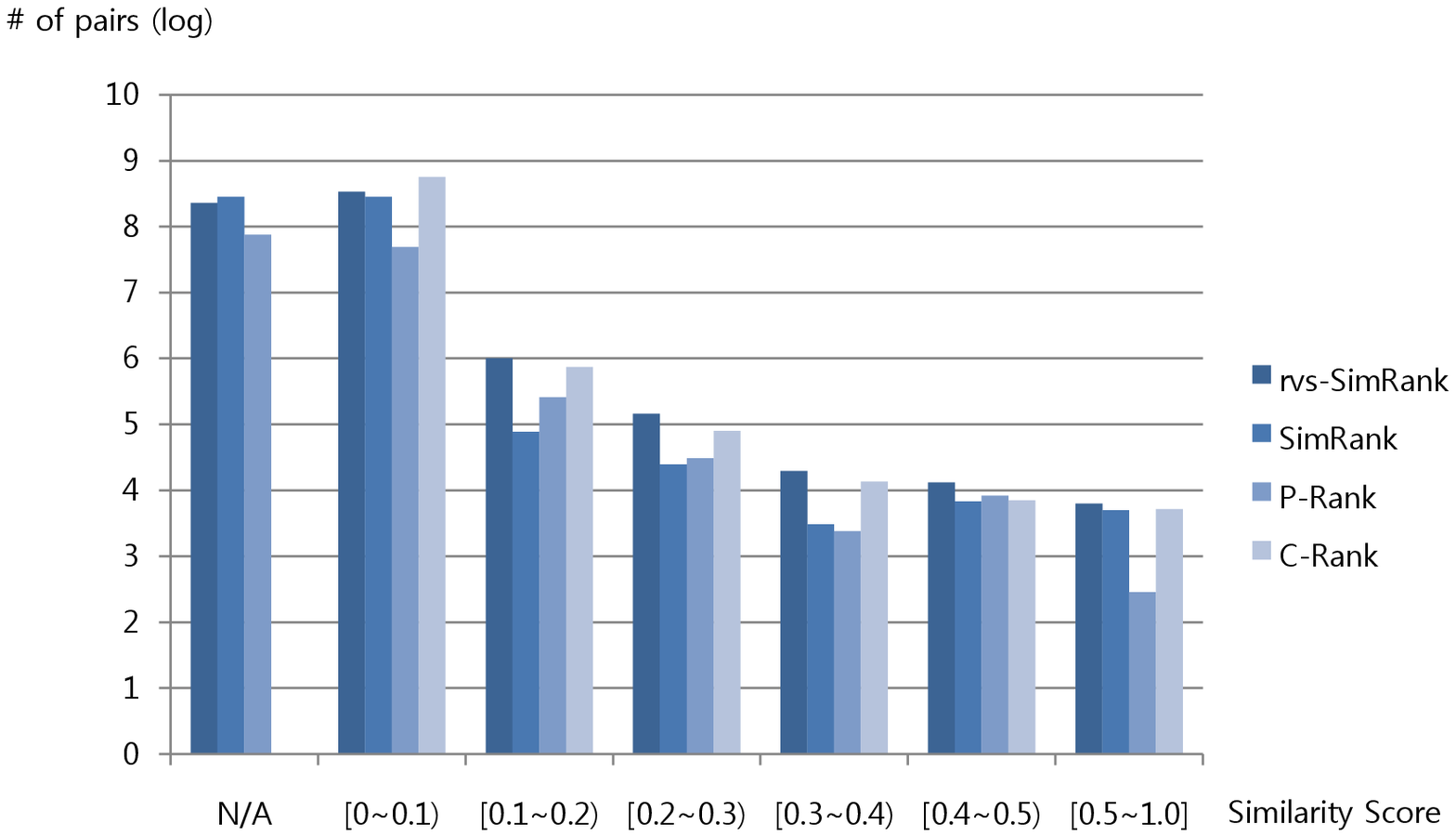,width=8cm} }
    \caption{Distributions of the similarity scores.}
    \label{fig:6}
\end{figure}

\subsubsection{Similarity Scores with Variations of the Number of Iterations}
In this section, we examine the algorithmic nature of similarity
measures by tracing the changes in the similarity score while
varying $k$. Figure 7 represents the average of the similarity
scores of the 10 highest-scoring pairs of papers while varying $k$
from 1 to 10. In Figure 7, x-axis represents the number of
iterations, and y-axis represents the average of the scores of the
top 10 highest-scoring pairs of papers by rvs-SimRank, SimRank,
P-Rank, and C-Rank, respectively. The similarity score $R_{k}(*,*)$
becomes more accurate on successive iterations. Iteration 2, which
computes $R_{2}(*,*)$ from $R_{1}(*,*)$, can be thought of as the
first iteration taking advantage of the recursive power of
algorithms for similarity computation. Subsequent changes become
increasingly minor, suggesting a rapid convergence. The score by
SimRank converges at $k=3$, the score by rvs-SimRank converges at
$k=5$, the score by P-Rank converges at $k=6$, and the score by
C-Rank converges at $k=9$. Because it utilizes the highest number of
links, C-Rank is the last one to converge.
\begin{figure}[h]
    \centerline{\psfig{figure=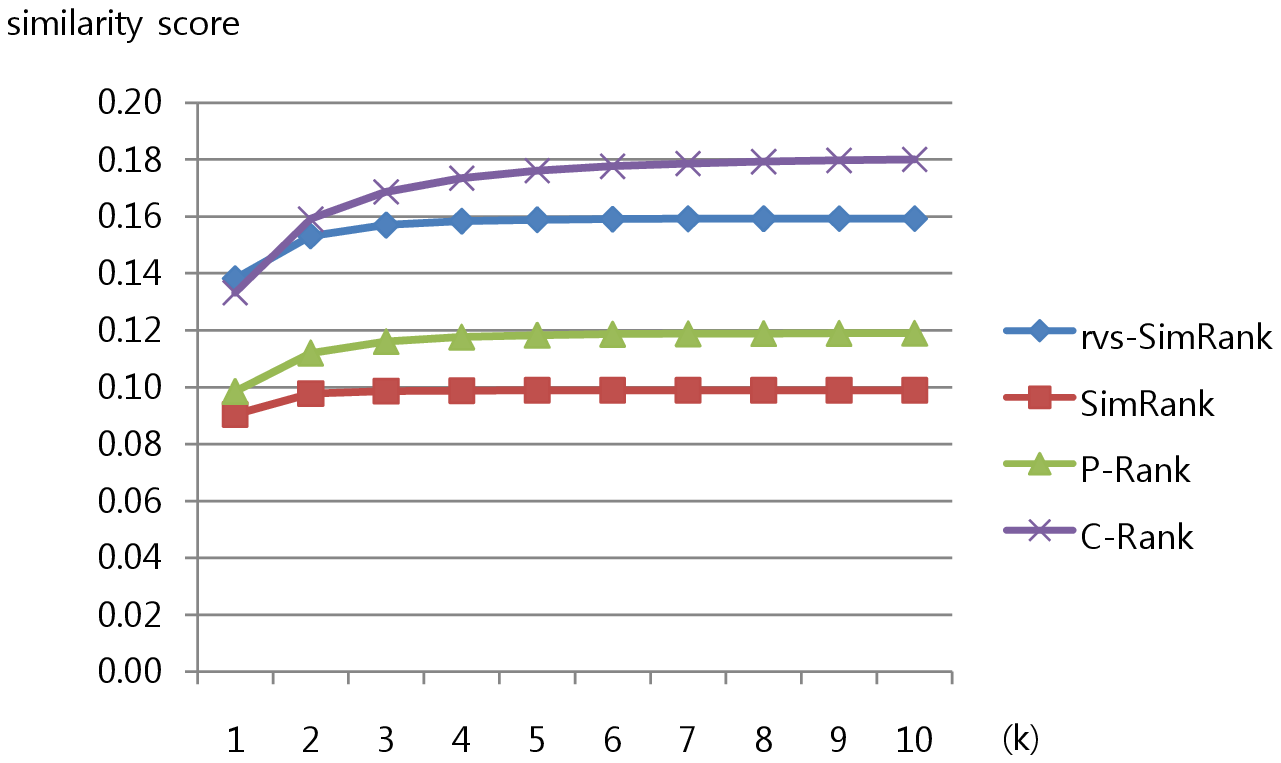,width=8cm} }
    \caption{The similarity scores with different $k$ values.}
    \label{fig:7}
\end{figure}

\subsubsection{Similarity Scores with Variations of Decay Factor}
In this section, we show how the decay factor $C$ is related to the
speed of convergence in C-Rank. Figure 8 represents the average
similarity scores by C-Rank with variations of $C$. In Figure 8,
x-axis represents the number of iterations, and y-axis represents
the average similarity score by the top 10 highest-scoring pairs of
papers. The decay factor, $C$, is set to be 0.2, 0.5, and 0.8,
respectively. It is obvious that the similarity score of C-Rank
increases with the increase of $C$. When $C=0.2$, C-Rank converges
fast at $k=2$. When $C=0.8$, on the other hand, C-Rank converges at
the $9$-th iteration. When $C$ is low, the recursive power of C-Rank
is weakened such that only the papers in local or near-local
neighborhood are used in similarity computation. When $C$ is high,
more papers in a more global neighborhood can be used in computing
the similarity recursively. When $C$ is high, therefore, the
convergence takes more time.

\subsubsection{Accuracy of Similarity Measures with Variations of the Relative Weight}

So far, we have used the relative weight to be 0.5 in P-Rank. In
this section, we compare the accuracy of C-Rank and those of P-Rank
with variations of $\lambda$. The $\lambda$ is set to be 0.3, 0.5,
and 0.8. Figure 9 represents the accuracy of C-Rank and P-Rank with
variations of $\lambda$. In Figure 9, x-axis represents the number
of the top $m$ scoring papers, and y-axis represents the accuracy of
each similarity measure. The accuracy of C-Rank is higher than those
of P-Rank regardless of the value of $\lambda$ in most cases.
Although the accuracy of P-Rank with $\lambda= 0.8$ is higher than
that of C-Rank in two cases, when $m=40$ and $m = 50$, the
similarity score is more important when $m$ is low, especially in
scientific literature retrieval services, and C-Rank achieves a
higher accuracy than P-Rank when $m$ is 10, 20, and 30.

\begin{figure}[h]
    \centerline{\psfig{figure=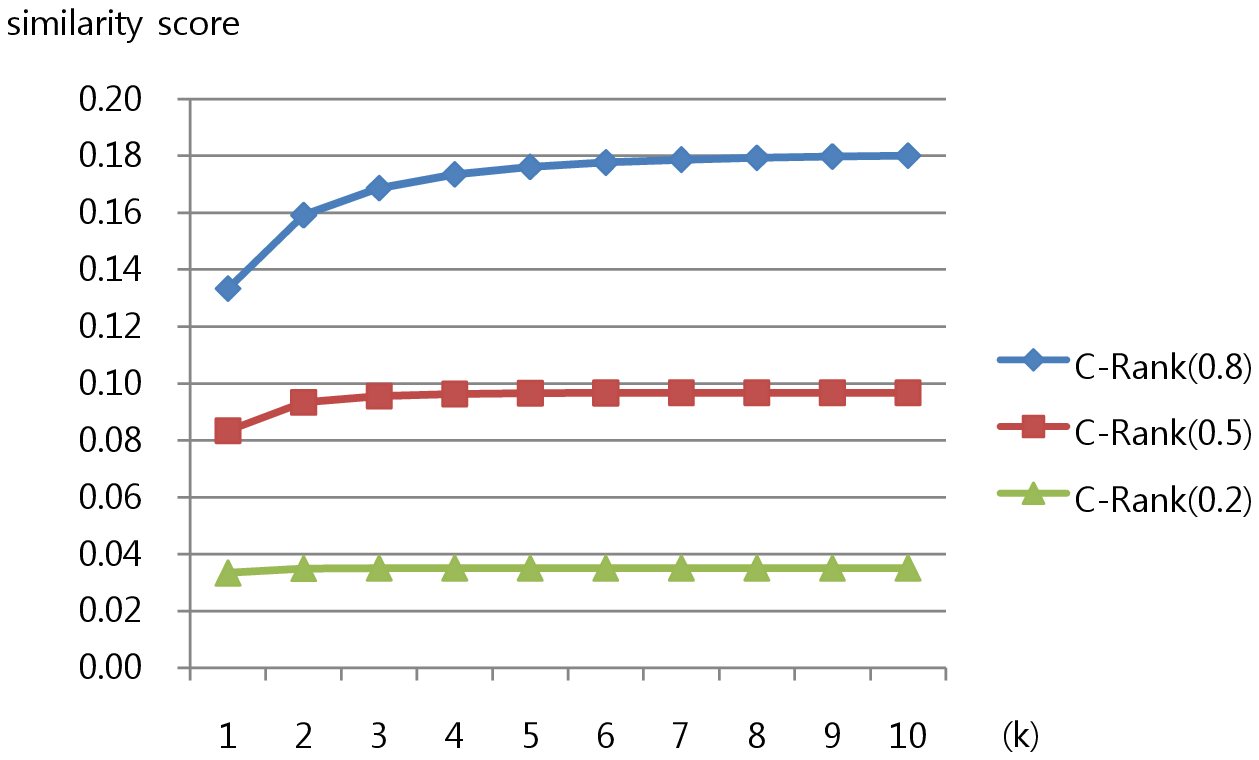,width=8cm} }
    \caption{The similarity scores with different $C$ values.}
    \label{fig:8}
\end{figure}

\begin{figure}[h]
    \centerline{\psfig{figure=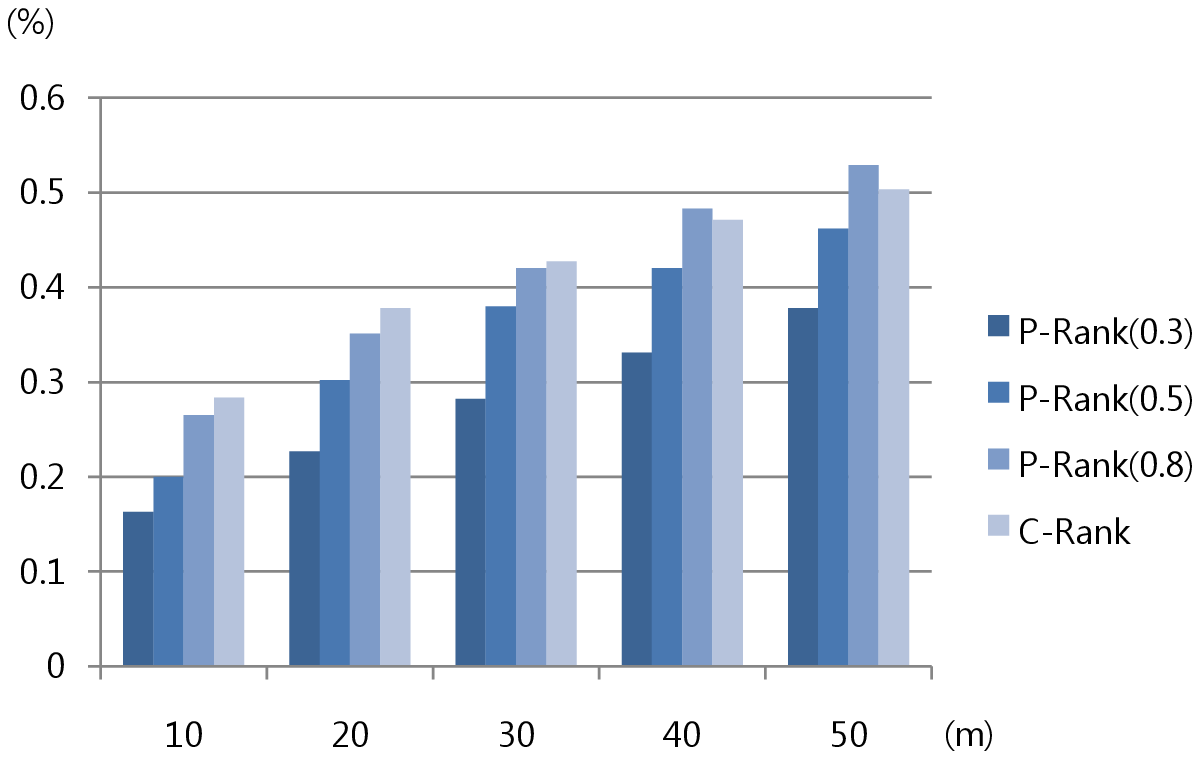,width=7cm} }
    \caption{The accuracy of C-Rank and P-Rank with different $\lambda$ values.}
    \label{fig:9}
\end{figure}

\section{Conclusions}
In this paper, we propose C-Rank, a new similarity measure for
scientific literature databases. We examine two notable
characteristics in scientific literature databases and identify
three cases where all existing similarity measures fail to compute
the similarity score correctly. Our observations lead to the
development of C-Rank, which uses both in-link and out-link while
disregarding the direction of references. In addition, we verify
Jaccard coefficient is more appropriate for scientific literature
databases, and propose an evaluation method for measuring the
accuracy of similarity measures. For experiments, we have built a
database with real papers from DBLP and reference information
crawled from Libra. Experimental results show that C-Rank achieves a
higher effectiveness than existing similarity measures in most
cases.

The contributions of this paper are as follows:
\begin{enumerate}
\item We have pointed out that existing similarity measures fail to
compute the similarity score properly for scientific papers.
\vspace{-0.2cm}
\item
We have proposed a new similarity measure for computing the
similarity score among papers called C-Rank. \vspace{-0.2cm}
\item We have proposed a normalization method suitable for scientific literature databases.
\vspace{-0.2cm}
\item We have proposed a quantitative evaluation method which matches the intuition of users.
\end{enumerate}

%

\vspace{1cm}
\section{Appendix}
{

We prove following four mathematical properties:

\renewcommand{\arraystretch}{1.05}
\begin{supertabular}{p{0.01cm}p{8.1cm}}\\
&\\
    1.& (Symmetry) According to Equation(8), it is $R_{k}(a,b) = R_{k}(b,a)$ for $k\geq 0$.\\
    &\\
    2.&(Monotonicity) If $a=b$, $R_{0}(a,b)=R_{1}(a,b)=...=1$, so it is that the monotonicity property holds. We consider $a\neq b$. According to Equation(8), $R_{0}(a,b)=0$. Base on Equation(8), $0\leq$ $R_{1}(a,b)\leq 1$. So, $0\leq R_{0}(a,b)\leq$ $R_{1}(a,b)\leq1$. We assume that for all $k$, $0\leq R_{k-1}(a,b)\leq R_{k}(a,b)\leq 1$, then \\
    &\\
    &$R_{k-1}(a,b)-R_{k}(a,b)=C\times$$\frac{|L(a)\setminus L(b)|}{|L(a)\cup L(b)|}\times$$\frac{1}{|L(a)\setminus L(b)||L(b)|}$\\
    &\\
    &$\times\sum\limits_{a' \in L(a) \setminus L(b)}\sum\limits_{b' \in L(b)}[R_{k}(a',b')-R_{k-1}(a',b')]$\\
    &\\
    &$+C\times\frac{|L(b)\setminus L(a)|}{|L(a)\cup L(b)|}\times\frac{1}{|L(b)\setminus
    L(a)||L(a)|}$\\
    &\\
    &$\times\sum\limits_{a' \in L(a) }\sum\limits_{b' \in L(b)\setminus
    L(a)}[R_{k}(a',b')-R_{k-1}(a',b')]$\\
    &\\
    &Based on the assumption, we have $(R_{k}(a,b)-R_{k-1}(a,b))\geq 0$, $\forall a,b\in G$, so the left hand side $R_{k+1}(a,b)-R_{k}(a,b)\geq 0$ holds. By induction, we draw the conclusion that for any $k$, $R_{k}\leq R_{k+1}$. And based on the assumption, $0\leq R_{k}(a,b)\leq 1$, so\\
    &\\
    &1) $C\times \frac{|L(a) \cap L(b)|}{|L(p) \cup I(a)|} \times 1$\\
    &\\
    &2) $C\times$$\frac{|L(a)\setminus L(b)|}{|L(a)\cup L(b)|}\times$$\frac{1}{|L(a)\setminus L(b)||L(b)|}$\\
    &\\
    &$\times\sum\limits_{a' \in L(a) \setminus L(b)}\sum\limits_{b' \in L(b)}R_{k}(a',b')$\\
    &\\
    &$\leq C\times \frac{|L(a) \setminus L(b)|}{|L(a) \cup L(b)|}\times\frac{1}{|L(a)\setminus L(b)||L(b)|}$\\
    &\\
    &$\times\sum\limits_{a' \in L(a) \setminus L(b)}\sum\limits_{b' \in L(b)}1=C\times \frac{|L(a)\setminus L(b)|}{|L(a)\cup
    L(b)|}$\\
    &\\
    &3) $C\times$$\frac{|L(b)\setminus L(a)|}{|L(a)\cup L(b)|}\times$$\frac{1}{|L(b)\setminus L(a)||L(a)|}$\\
    &\\
    &$\times\sum\limits_{a' \in L(a)}\sum\limits_{b' \in L(b)\setminus L(a)}R_{k}(a',b')$\\
    &\\
    &$\leq C\times \frac{|L(b) \setminus L(a)|}{|L(a) \cup L(b)|}\times\frac{1}{|L(b)\setminus L(a)||L(a)|}$\\
    &\\
    &$\times\sum\limits_{a' \in L(a)}\sum\limits_{b' \in L(b)\setminus L(a)}1$\\
    &\\
    &$=C\times \frac{|L(b)\setminus L(a)|}{|L(a)\cup L(b)|}$\\
    &\\
    &The above equation represents following\\
    &\\
    &$C[\frac{|L(a) \cap L(b)|}{|L(p) \cup I(a)|}+\frac{|L(a)\setminus L(b)|}{|L(a)\cup
    L(b)|}+\frac{|L(b)\setminus L(a)|}{|L(a)\cup L(b)|}]=C$\\
    &\\
    &so, $R_{k+1}(a,b) \leq C \leq 1$. By induction, we know that for any $k$, $0 \leq R_{k}(a,b) \leq 1$.\\
    &\\
    3.&(Existence) According to (Monotonicity), $\forall a,b \in G$, $R_{k}(a,b)$ is bounded and nondecreasing as $k$ increase. By the Completeness Axiom of calculus, each sequence $R_{k}(a,b)$ converges to a limit $R(a,b) \in [0,1]$. Note $\lim_{k \rightarrow \infty}R_{k}(a,b) = \lim_{k \rightarrow \infty}R_{k+1}(a,b) = R(a,b)$, So we have\\
    &\\
    &$R(a,b) = \lim\limits_{k \to \infty }R_{k + 1}(a,b)$ \\
    &\\
    &$=\lim\limits_{k \to \infty } C \times [\frac{{|L(a) \cap L(b)|}}{{|L(a) \cup L(b)|}} \times 1 + \frac{{|L(a)\setminus L(b)|}}{{|L(a) \cup L(b)|}}$\\
    &\\
    &$\times \frac{1}{{|L(a)\setminus L(b)||L(b)|}}\sum\limits_{a' \in L(a)\setminus L(b)} {\sum\limits_{b' \in L(b)} {{R_k}(a',b')} }  $\\
    &\\
    &$+ \frac{{|L(a)\setminus L(b)|}}{{|L(b) \cup L(a)|}} \times \frac{1}{{|L(b)\setminus L(a)||L(a)|}}\sum\limits_{a' \in L(a)}$\\
    &\\
    &$\times{\sum\limits_{b' \in L(b)\setminus L(a)}}{{R_k}(a',b')}]$\\
    &\\
    &$= C \times [\frac{{|L(a) \cap L(b)|}}{{|L(a) \cup L(b)|}} \times 1 + \frac{{|L(a)\setminus L(b)|}}{{|L(a) \cup L(b)|}}$\\
    &\\
    &$\times \frac{1}{{|L(a)\setminus L(b)||L(b)|}}\sum\limits_{a' \in L(a)\setminus L(b)} {\sum\limits_{b' \in L(b)} {\lim\limits_{k \to \infty } {R_k}(a',b')} }$  \\
    &\\
    &$+ \frac{{|L(a)\setminus L(b)|}}{{|L(b) \cup L(a)|}} \times \frac{1}{{|L(b)\setminus L(a)||L(a)|}}$\\
    &\\
    &$\times\sum\limits_{a' \in L(a)} {\sum\limits_{b' \in L(b)\setminus L(a)} }{\lim\limits_{k \to \infty }R_{k}(a',b')}]$ \\
    &\\
    &$= C \times [\frac{{|L(a) \cap L(b)|}}{{|L(a) \cup L(b)|}} \times 1 + \frac{{|L(a)\setminus L(b)|}}{{|L(a) \cup L(b)|}} $\\
    &\\
    &$\times \frac{1}{{|L(a)\setminus L(b)||L(b)|}}\sum\limits_{a' \in L(a)\setminus L(b)} {\sum\limits_{b' \in L(b)} {R(a',b')} }$  \\
    &\\
    &$+ \frac{{|L(b)\setminus L(a)|}}{{|L(a) \cup L(b)|}} \times \frac{1}{{|L(b)\setminus L(a)||L(a)|}}$\\
    &\\
    &$\times\sum\limits_{a' \in L(a)} {\sum\limits_{b' \in L(b)\setminus L(a)}}{R(a',b')}]  $\\
    &\\
    &Note that the limit of $R_{k}(*,*)$, with respect to $k$, right satisfies the recursive C-Rank equation, shown in Equation(8).\\
    &\\
    4.&(Uniqueness) Suppose $s_{1}(*,*)$ and $s_{2}(*,*)$ are two solution to the $n^{2}$ iterative C-Rank equations. for any entities $x,y \in G$, let $\delta(x,y) = s_{1}(x,y) - s_{2}(x,y)$ be their difference. Let $M = \max_{x,y}|\delta(x,y)|$ be the maximum absolute value of any difference. We need to show that $M = 0$. Let $|\delta(x,y)| = M$ for some $a,b \in G$. It is obvious that $M = 0$ if $a = b$. otherwise,\\
    &\\
    &$\delta(a,b) = s_{1}(a,b)-s_{2}(a,b)$\\
    &\\
    &$C \times\frac{|L(a)\setminus L(b)|}{L(a)\cup L(b)|} \times \frac{1}{|L(a)\setminus L(b)||L(b)|}$\\
    &\\
    &${\tiny\times\sum\limits_{a'\in L(a)\setminus L(b)}{\sum\limits_{b' \in L(b)}}[S_{1}(L(a'),L(b'))-S_{2}(L(a'),L(b'))]}$\\
    &\\
    &$+ C \times \frac{|L(b)\setminus L(a)|}{|L(a)\cup L(b)|} \times \frac{1}{|L(b)\setminus L(a)||L(a)|}$\\
    &\\
    &${\tiny\times\sum\limits_{a'\in L(a)}{\sum\limits_{b' \in L(b) \setminus L(a)}}[S_{1}(L(a'),L(b'))-S_{2}(L(a'),L(b'))]}$\\
    &\\
    &Thus,\\
    &\\
    &$M =|\delta(a,b)|= |C \times \frac{|L(a)\setminus L(b)|}{|L(a)\cup L(b)|}$\\
    &\\
    &$\times\frac{1}{|L(a)\setminus L(b)||L(b)|}\sum\limits_{a'\in L(a)\setminus L(b)}{\sum\limits_{b'\in L(b)}{\delta(a',b')}}$\\
    &\\
    &$+ C \times \frac{|L(b)\setminus L(a)|}{|L(a)\cup L(b)|}\times\frac{1}{|L(b)\setminus L(a)||L(a)|}$\\
    &\\
    &$\sum\limits_{a'\in L(a)}{\sum\limits_{b'\in L(b)\setminus L(a)}{\delta(a',b')}}|$\\
    &\\
    &$\leq|C \times \frac{|L(a)\setminus L(b)|}{|L(a)\cup L(b)|}\times\frac{1}{|L(a)\setminus
    L(b)||L(b)|}$\\
    &\\
    &$\times\sum\limits_{a'\in L(a)\setminus L(b)}{\sum\limits_{b'\in L(b)}{\delta(a',b')}}|$\\
    &\\
    &$+|C\times\frac{|L(b)\setminus L(a)|}{|L(a)\cup L(b)|}\times\frac{1}{|L(b)\setminus L(a)||L(a)|}$\\
    &\\
    &$\times\sum\limits_{a'\in L(a)}{\sum\limits_{b'\in L(b)\setminus L(a)}{\delta(a',b')}}|$\\
    &\\
    &$\leq C\times\frac{|L(a)\setminus L(b)|}{|L(a)\cup L(b)|}\times\frac{1}{|L(a)\setminus
    L(b)||L(b)|}$\\
    &\\
    &$\times\sum\limits_{a'\in L(a)\setminus L(b)}{\sum\limits_{b`\in L(b)}{|\delta(a',b')|}}$\\
    &\\
    &$+ C\times\frac{|L(b)\setminus L(a)|}{|L(a)\cup L(b)|}\times\frac{1}{|L(b)\setminus
    L(a)||L(a)|}\sum\limits_{a'\in L(a)}{\sum\limits_{b'\in L(b)\setminus L(a)}{|\delta(a',b')|}}$\\
    &\\
    &$\leq C\times\frac{|L(a)\setminus L(b)|}{|L(a)\cup L(b)|}$\\
    &\\
    &$\times\frac{1}{|L(a)\setminus
    L(b)||L(b)|}\sum\limits_{a'\in L(a)\setminus L(b)}{\sum\limits_{b`\in L(b)}{M}}$\\
    &\\
    &$+C\times\frac{|L(b)\setminus L(a)|}{|L(a)\cup L(b)|}$\\
    &\\
    &$\times\frac{1}{|L(b)\setminus
    L(a)||L(a)|}\sum\limits_{a'\in L(a)}{\sum\limits_{b'\in L(b)\setminus L(a)}{M}}$\\
    &\\
    &$\leq C\times\frac{|L(a)\setminus L(b)|}{|L(a)\cup L(b)|}\times M + C \times\frac{|L(b)\setminus L(a)|}{|L(a)\cup L(b)|}\times M$\\
    &\\
    &$\leq CM\times[\frac{|L(a)\setminus L(b)|}{|L(a)\cup L(b)|}+\frac{|L(b)\setminus L(a)|}{|L(a) \cup L(b)|}]$\\
    &\\
    &$\leq CM \qquad($by $\frac{|L(a)\setminus L(b)|}{|L(a) \cup L(b)|}+\frac{|L(b)\setminus L(a)|}{|L(a)\cup L(b)|}\leq1)$\\
    &\\
    &So $M = 0$ when $ C \neq1$.\\
\end{supertabular}
}

%
%
\end{document}